\documentclass[twocolumn,superscriptaddress,aps,longbibliography,showpacs,preprintnumbers,amsmath,amssymb,floatfix]{revtex4-1}
\usepackage{titlesec}
\usepackage{amsmath}
\usepackage{bm}
\usepackage{graphicx}
\usepackage[ansinew]{inputenc}
\usepackage{array}
\usepackage{color}
\usepackage{amsxtra}
\usepackage{amstext}
\usepackage{amssymb}
\usepackage{latexsym}
\usepackage{gensymb}
\usepackage{dsfont}
\usepackage{braket}
\usepackage[caption=false]{subfig}
\usepackage[makeroom]{cancel}

\usepackage{balance}

\usepackage{dcolumn}
\usepackage{bm}
\usepackage{bbm}
\usepackage{braket}
\usepackage{mathrsfs}
\usepackage{amstext}
\usepackage{euscript}
\usepackage{txfonts}

\usepackage[unicode=true,
 bookmarks=false,
 breaklinks=false,pdfborder={0 0 1},backref=false,colorlinks=true,citecolor=blue,urlcolor=blue,linkcolor=blue]
 {hyperref}

\def\B{\hat{B}}
\def\a{a_A}
\def\b{b_A}
\def\inpsi{\ket{\psi}_{in}}
\def\outpsi{\ket{\psi}_{out}}

\makeatother

\newcommand{\cref}[1]{Ref.\,\cite{#1}}

\newcommand*{\figref}[2][]{%
  \hyperref[{#2}]{%
    \ref*{#2}%
    \ifx\\#1\\%
    \else
      #1%
    \fi
  }%
}

\makeatother
\makeatletter
\renewcommand{\p@subsection}{}
\renewcommand{\p@subsubsection}{}
\makeatother

\begin{document}

\title{Optimal entanglement enhancing via conditional measurements}
\author{Jiru Liu}
\email{ljr1996@tamu.edu}
\affiliation{Institute for Quantum Science and Engineering (IQSE) and Department of Physics and Astronomy, Texas A\&M University, College Station, TX 77843-4242, USA}

\author{Yusef Maleki}
\email{maleki@tamu.edu}
\affiliation{Institute for Quantum Science and Engineering (IQSE) and Department of Physics and Astronomy, Texas A\&M University, College Station, TX 77843-4242, USA}

\author{M. Suhail Zubairy}
\email{zubairy@physics.tamu.edu}
\affiliation{Institute for Quantum Science and Engineering (IQSE) and Department of Physics and Astronomy, Texas A\&M University, College Station, Texas 77843-4242, USA}
\date{\today}

\begin{abstract}
Enhancing quantum entanglement is important for many quantum information processing applications. In this paper, we consider a protocol for entanglement enhancing in a two-mode squeezed vacuum state (TMSVS), attained based on photon subtraction, photon catalysis, and photon addition.
Central to such an operation is the task of mixing and detecting number states with each mode of TMSVS. We analyze various settings and find an optimal setup for improving the entanglement of the state.
\end{abstract}
\maketitle

\section*{\centering\uppercase\expandafter{\romannumeral1}. Introduction}

Among various entangled states, continuous variable (CV) systems have attracted considerable attention for their remarkable characteristics, and their usefulness in quantum information tasks \cite{braunstein2005quantum,lloyd1999quantum,adesso2014continuous,lund2014boson,duan2000inseparability}. The prominent examples of CV systems include both Gaussian and non-Gaussian states. It is well-known that Gaussian states, such as coherent and squeezed states,  offer a considerable platform for quantum applications. In a parallel line of research studies, it has been shown that non-Gaussian states, as well as non-Gaussian operations, can play a significant role in quantum information processing. For example,  two-mode non-Gaussian states present an advantage over Gaussian states in enhancing entanglement \cite{lee2011enhancing,navarrete2012enhancing,walschaers2017statistical}. This is an interesting property due to the fact that highly entangled states are of particular importance for both practical applications \cite{horodecki2009quantum,renner2008security,pirandola2015advances,zubairy2020quantum,maleki2021quantum,ge2020operational,pirandola2015advances,zubairy2020quantum,maleki2021quantum,ge2020operational,barnett2009quantum} and fundamental investigations in quantum discipline \cite{barnett2009quantum,maleki2021quantum,genovese2005research}. Moreover, non-Gaussian sources are necessary for entanglement distillation since distilling Gaussian states via Gaussian operations is not possible \cite{giedke2002characterization,fiuravsek2002gaussian}.     
\par
Non-Gaussian states can be obtained by simply adding or subtracting photons to Gaussian states. Photon addition and photon subtraction are important tools for improving quantum correlations. The physical properties of photon-added and photon-subtracted non-Gaussian states are studied both theoretically and experimentally, in recent years \cite{agarwal1991nonclassical,ourjoumtsev2009preparation,walschaers2017statistical,zavatta2004quantum,ourjoumtsev2007increasing}. Photon addition and photon subtraction on squeezing states and coherent state are used for entanglement distillation \cite{lee2011enhancing,zhang2013continuous}, quantum commutation applications \cite{zavatta2009experimental,parigi2007probing}, entanglement and the teleportation fidelity enhancements \cite{opatrny2000improvement,walschaers2017statistical,hu2017continuous,ourjoumtsev2007increasing,yang2009entanglement,lee2011enhancing,navarrete2012enhancing,zhang2011local,fiuravsek2011improving,dell2010realistic,dell2007continuous}. 
\par 
Photon subtraction is shown to be realized by taking a small fraction out of the light beam \cite{wenger2004non,opatrny2000improvement}. Also,  photon addition is demonstrated to be attained in parametric down-conversion processes in BBO crystal \cite{zavatta2004quantum}. Both experimental methods include conditional measurements. Considering these approaches, a similar non-Gaussian operation, photon catalysis, was also studied by A. I. Lvovsky and J. Mlynek \cite{lvovsky2002quantum}. Photon catalysis has been demonstrated to enhance the entanglement of a two-mode squeezed vacuum state \cite{hu2017continuous}.
\par
The importance of the non-Gaussian entanglement enhancement recipes that were mentioned earlier becomes more transparent by considering the fact that generating highly entangled states is not an easy task in general and requires a delicate control and design of quantum system \cite{zubairy2020quantum,barnett2009quantum,scully1999quantum}.
 For example, two-mode squeezed vacuum state (TMSVS), can be attained using non-linear crystals \cite{taylor2016quantum}. However, due to the weak interaction of non-linear processes, the squeezing factor is usually small. To be more precise,  the entanglement of a TMSVS is determined by $E_\mathcal{N}=\log_2e^{2r}$ where $E_\mathcal{N}$ is the logarithmic negativity \cite{agarwal2012quantum}, and $r$ is the squeezing factor.
\par
In this study, we consider the problem of enhancing the entanglement in TMSVS based on schemes that take advantage of the photon addition, subtraction, and catalysis phenomena.
We first analyze and compare the performance of all three operations \cite{hu2017continuous,xu2015enhancing}. As demonstrated in Fig. \ref{fig:single_BS}, an auxiliary Fock state $\ket{m}_A$ is mixed with one of the modes of a TMSVS via a BS (see Fig. \ref{fig:single_BS}). Using time-multiplexed photon number resolving detector \cite{mivcuda2008high,fitch2003photon}, conditional measurement can be performed on the auxiliary mode of the output states, projecting it to $\ket{m'}_A$. If $m'<m$, this process adds photons to the states, while $m'>m$ subtracts photons from the state. For $m'=m$ it serves as a catalyst. 
The operation can be performed on both modes of the TMSVS as depicted in Fig. \ref{fig:two_BS}, where each mode gets mixed with an independent Fock state.
We find that photon catalysis provides better results for both success probability and entanglement enhancement compared to the photon addition and subtraction processes. However, even for the photon catalysis, the success probability of the states does not exceed 20\%.
\par
To overcome this limitation and to attain even considerably higher entanglement enhancement, we introduce a different method compared to previous studies  \cite{lee2011enhancing,navarrete2012enhancing,opatrny2000improvement,zhang2011local,hu2017continuous,xu2015enhancing}. We show that by injecting the auxiliary Fock states into a BS before mixing them with the modes of the TMSVS, the entanglement can be drastically improved. With the new setup, we demonstrate that under photon addition, the success probability for a large entanglement enhancement can even exceed 70\%. Our investigations show that adding a single photon to the TMSVS, based on this protocol, is, in fact, optimal for entanglement enhancement.

\section*{\centering\uppercase\expandafter{\romannumeral2}.
Photon subtraction, addition, and catalysis}

\begin{figure} 
\begin{center}

\includegraphics[width=7cm]{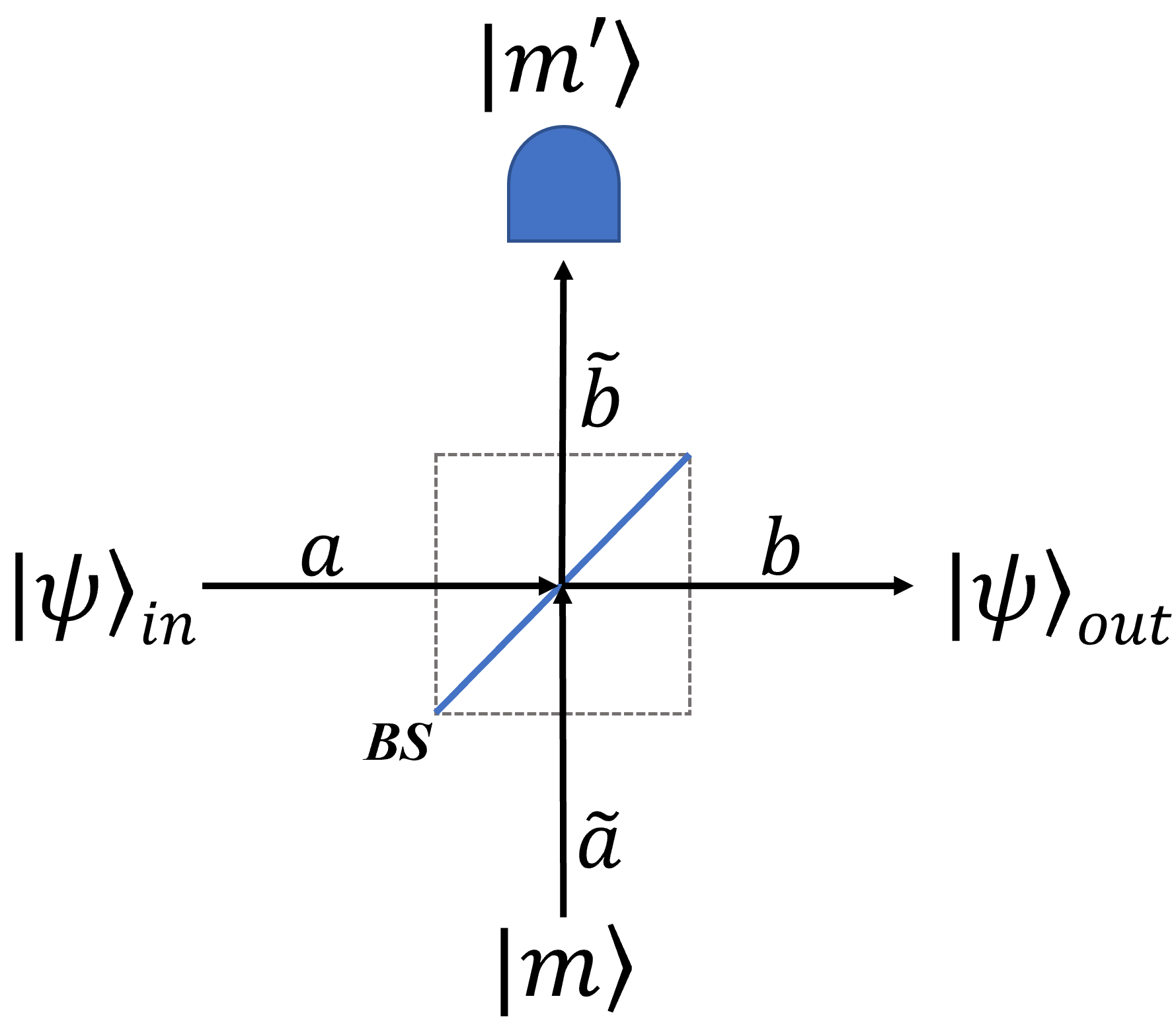}
\caption{Quantum state $\ket{\psi}_{out}$ is generated by mixing the Fock state $\ket{m}$ and the input state $\ket{\psi}_{in}$ in the BS, where the state $\ket{m'}$ is detected in the output mode.}
\label{fig:single_BS}
\end{center}
\end{figure}

Now, we consider a general setting of non-Gaussian operations, where photons can be added, subtracted, or catalyzed in the process.  
The protocol for enhancing entanglement is depicted in Fig. \ref{fig:single_BS}, where an input state $\ket{\psi}_{in}$ is injected into one port of a BS and the Fock state $\ket{m}$ into the other port. The BS has the transmittance $T=\cos^2\theta$. The detector in the output mode detects ${m'}$ photons, which can be realized using a time-multiplexed photon number resolving detector \cite{mivcuda2008high,fitch2003photon}. 
\par 
Based on the configurations in Fig. \ref{fig:single_BS}, there are three possible scenarios: Photon subtraction for $m'>m$, photon addition for $m'>m$, and photon catalysis $m'=m$. 
 The BS operation can be described as $\hat{B}(\theta)=\exp\{\theta(a^\dag a_A-{a_A}^\dag a)\}$, where for convenience we take $\theta$ to be real and the phase shift of BS is set to be zero.  Together with the photon number measurement operation, the output state $\ket{\psi}_{out}$ is given by
\begin{equation} \label{eq:psi}
    \ket{\psi}_{out}= _A\bra{m'}\hat{B}(\theta)\ket{m}_A\ket{\psi}_{in},
\end{equation}
where index $A$ denotes the auxiliary mode. Note that BS operator consists of two modes and hence ${}_A\bra{m'}\hat{B}(\theta)\ket{m}_A$ is an operator acting on the mode that is not measured.
Thus, the transformation from $\ket{\psi}_{in}$ to $\ket{\psi}_{out}$ is expressed by 
\begin{equation}
\label{eq:Bmm'}
\begin{aligned}
    \ket{\psi}_{out}&=\hat{B}_{m,m'}\ket{\psi}_{in},\\
    \hat{B}_{m,m'} &= {}_A\bra{m'}\hat{B}(\theta)\ket{m}_A.
\end{aligned}
\end{equation}

Considering Eq. (\ref{eq:Bmm'}), in order to determine the output state, we need to determine $\hat{B}_{m,m'}$. As is shown in the Appendix, the operator $\hat{B}_{m,m'}$ can be expressed in the Fock basis through
\begin{equation} \label{eq:Bm1_}
\begin{aligned}
\hat{B}_{m,m'} = _A\bra{m'}\hat{B}(\theta)\ket{m}_A =\sum_{k=0}B_{m,m',k}\ket{k+m-m'}\bra{k},
\end{aligned}
\end{equation}
where the explicit expression for the coefficient $B_{m,m',k}$ is presented in Eq. (\ref{A3}), in the Appendix.
This provides a rather general framework for non-Gaussian operations on a quantum state. 

\par 
To consider a particular setting, for the input state $\ket{\psi}_{in}=\sum_{k}c_k\ket{k}$, the output state $\outpsi$ can be obtained to be 
\begin{equation}
\label{eq:psi_out}
\ket{\psi}_{out}\propto \hat{B}_{m,m'}\inpsi=\sum_{k=0}c_kB_{m,m',k}\ket{k+m-m'}.
\end{equation}
Note that $\hat{B}_{m,m'}\inpsi$ is not normalized in general. Therefore, considering the normalized factor $N_{m,m'}$, determined by $N^{-2}_{m,m'}=\sum_k|c_kB_{m,m',k}|^2$, the output state can be expressed as 
\begin{equation} \label{eq:psiout5}
    \outpsi=N_{m,m'}\sum_{k=0}c_kB_{m,m',k}\ket{k+m-m'},
\end{equation}
\par 
As a particular example, for the photon catalysis with $m=m'=1$, we find for the output state
\begin{equation}
\begin{aligned}
     \outpsi&=N_{1,1}\sum_{k=0}c_k(\cos\theta)^{k-1}[\cos^2\theta-k\sin^2\theta] \ket{k},
\end{aligned}
\end{equation}
which agrees with Refs.\cite{hu2017continuous,sanaka2006filtering,resch2007entanglement}.
Taking into account the fact that $T=\cos^2\theta$ and $R=\sin^2\theta$, the equation above can be expressed as $ \outpsi=N_{1,1}\sum_{k=0}c_k\sqrt{T}^{k-1}[T-k R] \ket{k}$. Therefore, for $k=T/R$, the contribution from the Fock state $\ket{k}$ becomes eliminated in the output state.
As an specific case, for $ \inpsi=\ket{k}$, we obtain $\outpsi=0$.
For a given state, if we take $T=R$, for instance, there will be no contribution from the number state $ \ket{1}$, in the output. While, for $T=2R$ there will be no contribution from the number state $ \ket{2}$
in the output state, for instance.

\section*{\centering\uppercase\expandafter{\romannumeral3}. Entanglement enhancement of the TMSVS}

Now that we have developed a platform for desired non-Gaussian operations, we can 
implement specific setups for enhancing  entanglement in a continuous variable system. Even though the input state introduced in Fig. \ref{fig:single_BS}, is quite general, we apply the quantum enhancement protocols to the two-mode squeezed vacuum state (TMSVS).
The two-mode squeezed vacuum state  is defined by applying two-mode squeezing operator $S(r)=\exp\{r(a_1^\dag a_2^\dag-a_1a_2)\}$ to the two-mode vacuum state  $\ket{0,0}$ as \cite{hu2017continuous,sanaka2006filtering,resch2007entanglement}
\begin{equation} \label{eq:TMSV}
\begin{aligned}
    \ket{\text{TMSVS}}=&S(r)\ket{0,0}
    =\text{sech}r\sum_{k=0}\lambda^k \ket{k,k},
\end{aligned}
\end{equation}
where $\lambda=\tanh{r}$ (for convinence, $r$ is set to be real). The operators $a_1^\dag (a_1)$ and $a_2^\dag (a_2)$ are the creation (annihilation) operators for the two modes.

In particular, we consider two different setups for entanglement enhancements and determine the optimal scenario to achieve the goal.

\subsection*{A. First setup}

\begin{figure}[h] 
\begin{center}
\includegraphics[width=8cm]{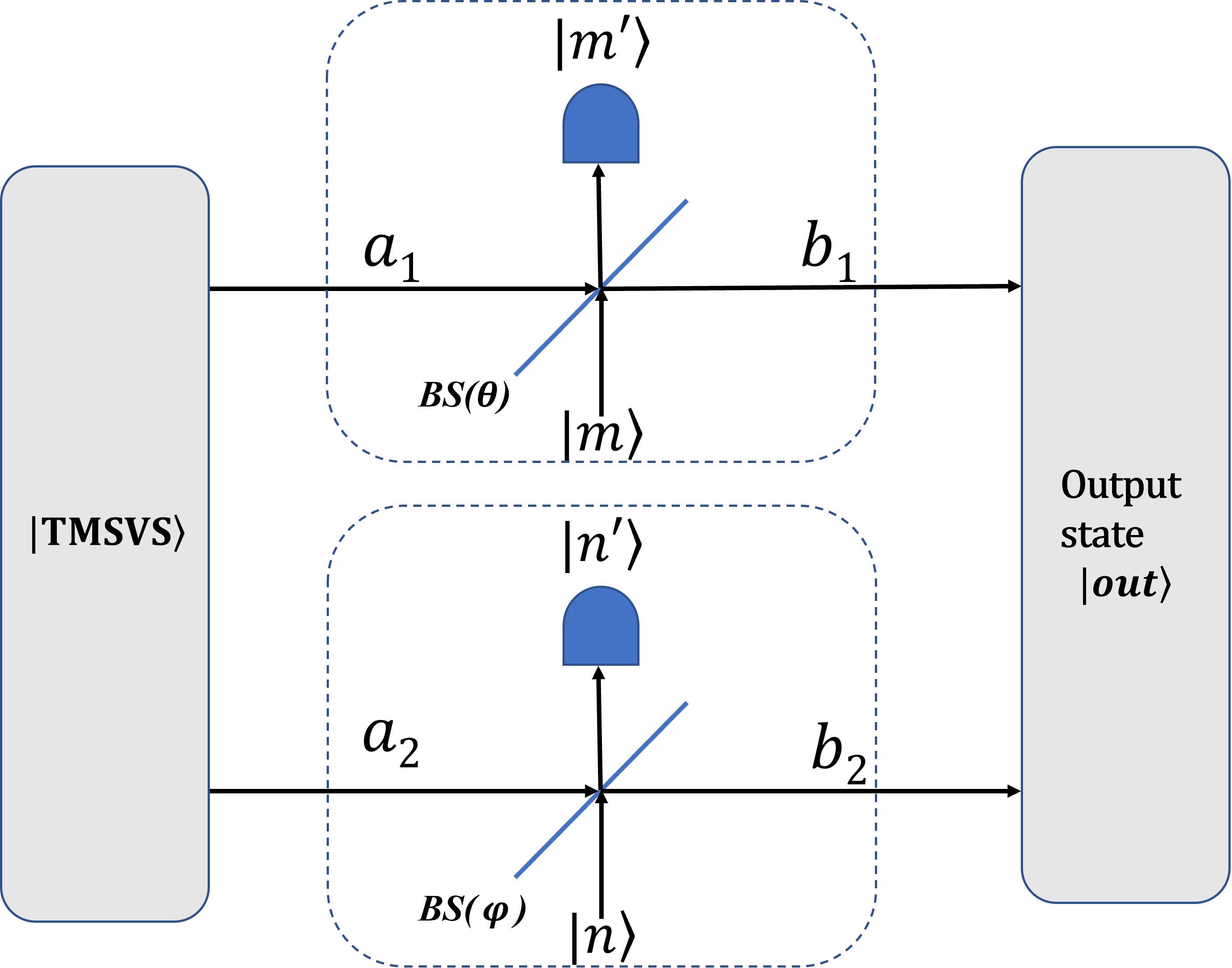}
\caption{The two modes of TMSVS are injected into two BSs separately, generating a two-mode output state $\ket{out}$.}
\label{fig:two_BS}
\end{center}
\end{figure}

Now, we introduce the first scheme for the enhancement of the entanglement in TMSVS. The setup is presented in Fig. \ref{fig:two_BS}, where each mode goes though the operation that is explained in Fig. \ref{fig:single_BS}. Considering the formalism developed in the previous section, the output state is given by
\begin{equation} \label{eq:out}
\begin{aligned}
    \ket{out}&=N_{m,m':n,n'}\B_{m,m'}(\theta)\B_{n,n'}(\varphi)\ket{\text{TMSVS}}\\
    &=N_{m,m':n,n'}\sum_{k=0}C_{m,m':n,n',k}\ket{k+m-m',k+n-n'},
\end{aligned}
\end{equation}
where $C_{m,m':n,n',k}$ is the coefficient of $\ket{k+m-m',k+n-n'}$. Note that $N_{m,m':n,n'}$ is the normalization factor such that $\bra{out}out\rangle=1$. $C_{m,m':n,n',k}$, which we denote as $C_k$ for convenience, can be obtained using the two separate non-Gaussian operations operated on each mode. In this setting, we have $C_k=\text{sech}r\ \lambda^k\ B_{m,m',k}(\theta)\ B_{n,n',k}(\varphi)$.

In this operation, the success probability and the entanglement enhancement are the two quantities that determine how useful the operation outcome is. $N^{-2}_{m,m':n,n'}$ quantifies the success probability of the conditional measurement $m\rightarrow m'\ \text{and}\ \ n\rightarrow n'$, which is determined by $N^{-2}_{m,m':n,n'}=\sum_{k=0}|C_k|^2$. For quantifying entanglement, we use the logarithmic negativity $E_\mathcal{N}$ as a measure of entanglement which is given by \cite{agarwal2012quantum}
\begin{equation} \label{logar}
    E_\mathcal{N}(\rho)=\log_2||\rho^{T_A}||_1,
\end{equation}
where $||R||_1$ denotes the trace norm $\text{Tr}\sqrt{R^{\dagger}R}$ and $\rho^{T_A}$ is the partial transpose of the state $\rho$. In Eq. (\ref{eq:out}), $\ket{out}$ is in a Schmidt form, for which the logarithmic negativity is given by
\begin{equation} \label{eq:logar2}
    E_\mathcal{N}=\log_2\left[N^2_{m,m':n,n'}(\sum_{k=0}|C_k|)^2\right].
\end{equation}
This formulation clearly shows that the entanglement is determined by the coefficient norms $|C_k|$.  Considering the explicit form of the coefficients as $C_k=\text{sech}r\ \lambda^k\ B_{m,m',k}(\theta)\ B_{n,n',k}(\varphi)$
it is evident that the characteristics of the beam-splitters and also $m,m',n,$ and $n'$ determine the entanglement. In general, the entanglement can be increased or decreased for some specific parameters. However, the interesting scenario is to find the parameter space where the entanglement increases. 

\begin{figure}[htbp]
\begin{center}
\includegraphics[width=9cm]{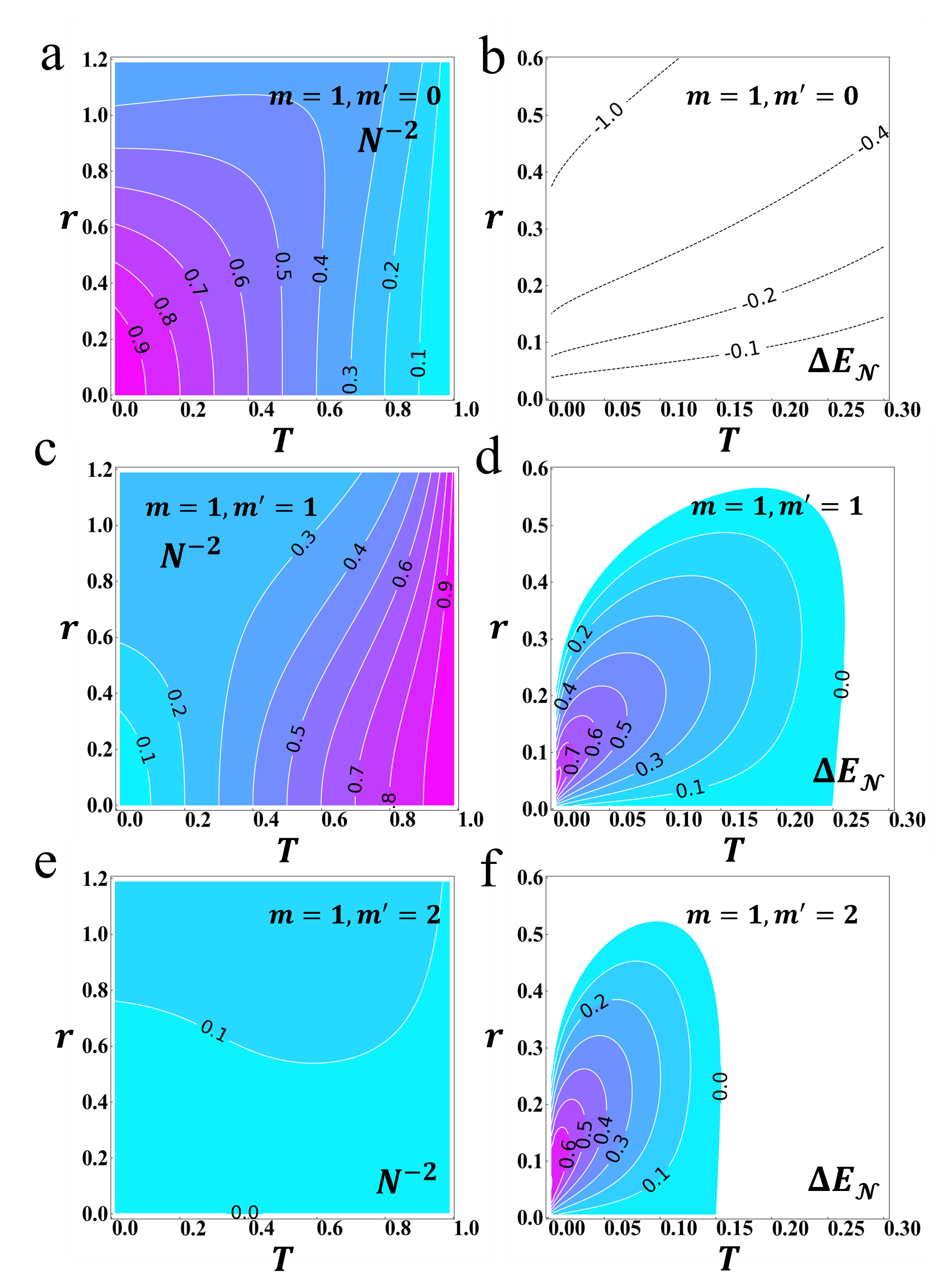}
\caption{The three operations for $m=1$. (a) and (b) for photon addition. (c) and (d) for photon catalysis. (e) and (f) for photon subtraction. (a),(c) and (e) present the success probability versus the squeezing factor $r$ and the transmittance $T=\cos^2\theta$, while (b), (d) and (f) show the entanglement enhancement for different cases.}
\label{fig:three}
\end{center}
\end{figure}

The initial entanglement for TMSVS, whose Schmidt form is shown by Eq. (\ref{eq:TMSV}), can be calculated from Eq. (\ref{eq:logar2}) that results in $E_\mathcal{N}(\text{TMSVS})=2r\log_2e.$ Therefore, the entanglement change is determined by
\begin{equation} \label{eq:DeltaE}
    \Delta E_\mathcal{N}=\log_2\left[N^2_{m,m':n,n'}(\sum_{k=0}|C_k|)^2\right]-2r\log_2e.
\end{equation}
\par
This relation is central to the investigation of the entanglement improvement by the protocol. In fact, if $E_\mathcal{N}$ is positive, then we can conclude that the entanglement has increased. To find the parameter space in which entanglement can be enhanced, we consider various settings and identify the cases that entanglement can be enhanced more efficiently. To be more specific, we compare the enhancement for the photon addition, subtraction and catalysis
and determine which operation can increase the entanglement better.

In some of the recent studies, photon subtraction (addition) are implemented by directly applying an annihilation (creation) operator to the input state, the theoretical analysis are based on the normalized state $\hat{a}\ket{\psi}_{in}$, $\hat{a}^\dag\ket{\psi}_{in}$ and the coherent superposition $(t\hat{a}+r\hat{a}^\dag)\ket{\psi}_{in}$ \cite{navarrete2012enhancing,lee2011enhancing,lee2010quantum,agarwal1991nonclassical,biswas2007nonclassicality}. Our scheme in Fig. \ref{fig:single_BS}  adds or subtracts photons or operates photon catalysis by a beam-splitter and conditional measurement. The analysis and simulation are based on the output state $\ket{\psi}_{out}$ given by Eq. (\ref{eq:psiout5}).
\par
We plot the outcome for the three operations for the case with $m=1$ in Fig. \ref{fig:three}. In this figure, for convenience and without loss of generality, no operations are applied to the lower mode of the TMSVS. For photon addition (Figs. \figref[(a)]{fig:three} and \figref[(b)]{fig:three}) $m'=0$, one photon is added to the TMSVS. As is shown by Fig. \figref[(b)]{fig:three}, there is no entanglement enhancement in the whole parameter space, in this case. Therefore, such a photon addition is not useful for entanglement improvements. For photon subtraction (Figs. \figref[(e)]{fig:three} and \figref[(f)]{fig:three}) $m'=2$, one photon is subtracted from the TMSVS. As we can see from Fig. \figref[(e)]{fig:three}, this operation has a quite small success probability in the regions where entanglement can be enhanced.
\par
Photon catalysis with $m'=1$ is shown in Figs. \figref[(c)]{fig:three} and \figref[(d)]{fig:three}. These plots show that the entanglement can be enhanced using a catalysis setting. However, the entanglement enhancement regions in Fig. \figref[(d)]{fig:three} correspond to the regions with the low success probability in Fig. \figref[(c)]{fig:three}. Nevertheless, the entanglement enhancement, in this case, is better than both the photon addition and photon subtraction settings \cite{hu2017continuous}.
It is worth mentioning that, even though we presented the $m=1$ scenario in Fig. \ref{fig:three}, this finding is true beyond this specific case, where one can consider larger $m$ and a various number of photon detection in the output mode. Therefore, our study unifies the previous consideration in the literature in a comparative setting and shows how various proposals can be compared with each other. Even though the observation that photon catalysis is a better route to quantum entanglement enhancement compared to the two other cases, in the  Fig. \ref{fig:two_BS} framework, the outcome still is not very compelling due to the low success probabilities. To overcome this limitation, we propose a different protocol enabling much higher success probabilities and large entanglement enhancements in the following subsection.

\subsection*{B. Second setup}
Now, we introduce a method for increasing entanglement that is different from the one depicted in Fig. \ref{fig:two_BS}.  We already considered the three non-Gaussian operations on the TMSVS where each mode undergoes a separate operation,  using the product Fock states $\ket{m}\otimes\ket{n}$ as auxiliary states. Among these schemes, we found that photon catalysis has the best success probability as well as entanglement improvement compared to photon addition and subtraction. However, according to Figs. \figref[(c)]{fig:three} and \figref[(d)]{fig:three}, for a suitable $r$ and $T$, the region which corresponds to a considerable entanglement enhancement has a moderately low success probability that is approximately less than 20\%. 
\par
\begin{figure}[htbp]
\begin{center}
\includegraphics[width=7.5cm]{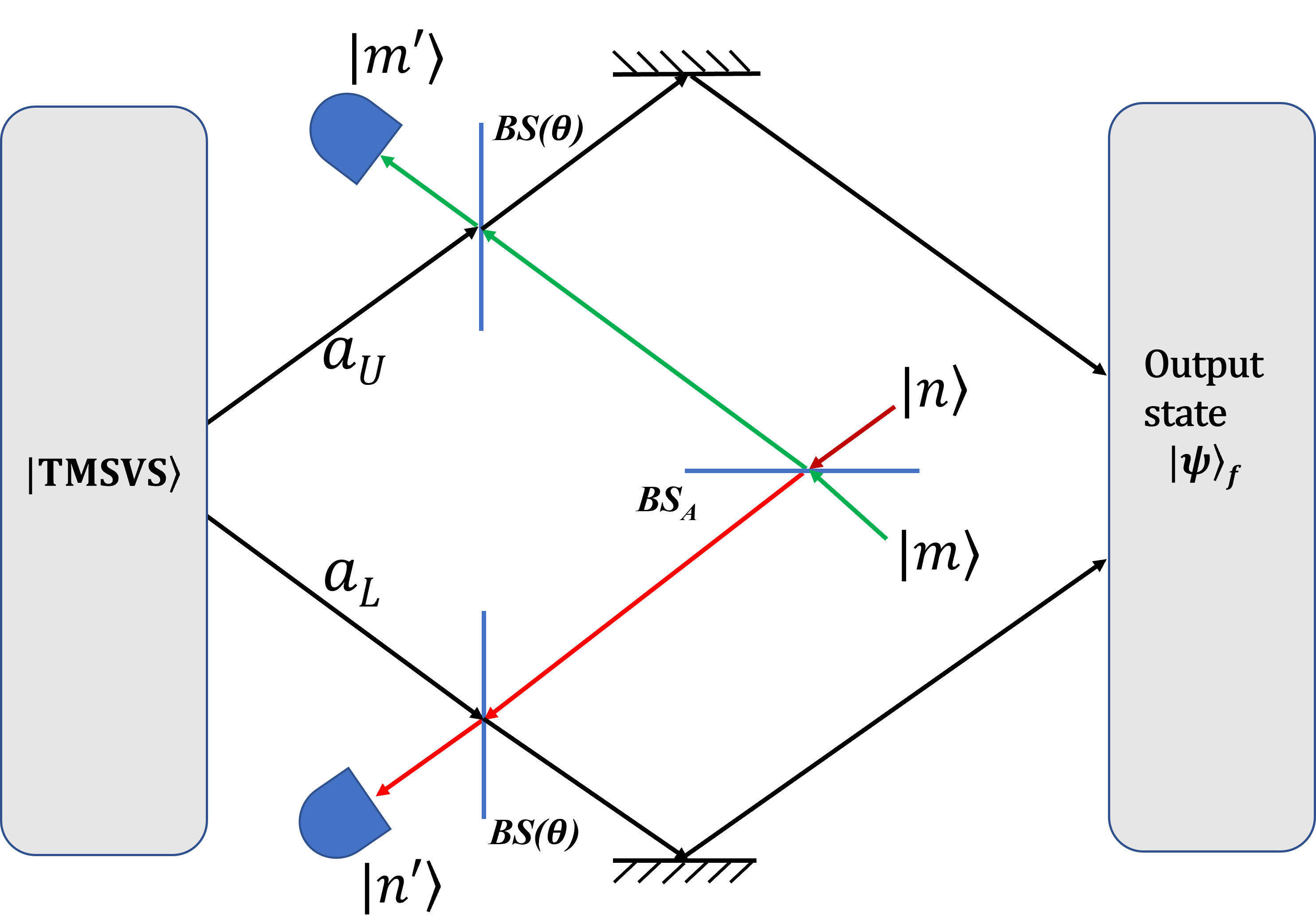}
\caption{A similar setting to Fig. \ref{fig:two_BS}, but the auxiliary states are first mixed via $BS_A$, before the operations are carried out on TMSVS.}
\label{fig:threeBS}
\end{center}
\end{figure}

We show that a slight adjustment in the setup can drastically enhance both the entanglement and the success probability. This adjustment requires an extra BS ($BS_A$), as is shown in Fig. \ref{fig:threeBS}. In this setting, the auxiliary input states for each modes first undergo $BS_A$ before the operation of Fig. \ref{fig:two_BS} is performed. This process can entangle the two auxiliary modes, and finally apply the non-Gaussian operations to the modes of TMSVS \cite{biagi2020entangling,ourjoumtsev2007increasing}. Since this extra step is unitary, no photon is lost in this process. This extra step is not challenging to implement in the experiment due to the simplicity of BS operations in general. Therefore, any improvement attained in the process can be useful in a practical setting.
\par
In Fig. \ref{fig:threeBS}, we simplify the auxiliary input source to one single photon state ($m=1,n=0$). Once $\ket{\psi}_A$ is mixed with the TMSVS via $BS(\theta)$, the conditional detection of $\ket{00}_A$ ($m'=0,n'=0$) adds one photon to the TMSVS.
\par
For $m=1, n=0$, before $BS_A$, the entire state is
\begin{equation} \label{eq:psiin_12}
\begin{aligned}
    \ket{\psi}_i&=\text{sech}r\sum_{k=0}\lambda^k\ket{k}_U\ket{k}_L\ket{1}_{UA}\ket{0}_{LA}\\
    &=\text{sech}r\sum_{k=0}\lambda^k\ a_{UA}^\dag\ \ket{kk}\ket{00}_A\\
    &=\text{sech}r\sum_{k=0}\frac{\lambda^k}{k!}\ (a_U^\dag)^k\ (a_{UA}^\dag)\ (a_L^\dag)^k\ \ket{00}\ket{00}_A,
\end{aligned}
\end{equation}
where the creation operator of the upper TMSVS mode is denoted as $a^\dag_U$ while the lower one is denoted as $a^\dag_L$. Moreover, the creation operator for the upper (shown in green) path of the ancillary state in Fig. \ref{fig:threeBS} is denoted as $a^\dag_{UA}$ while the lower (shown in red) one is $a^\dag_{LA}$. In the second step of above equations, $\ket{ij}_A$ stands for $\ket{i}_{UA}\ket{j}_{LA}$, and $\ket{ij}$ is simply the basis for the two modes of TMSVS, $\ket{i}_{U}\ket{j}_{L}$. 
\par
After passing through BS$_A$, we attain $\ket{\psi}_1$. This state, when it passes through BS$_U$ and BS$_L$, degenerates to $\ket{\psi}_2$, where, $\ket{\psi}_1$ and $\ket{\psi}_2$ are given by
\begin{equation} \label{eq:two_BS_state}
\begin{aligned}
    \ket{\psi}_1&=\frac1{\sqrt{2}}\text{sech}r\sum_{k=0}\frac{\lambda^k}{k!}\ (a_U^\dag)^k\ (a_{UA}^\dag+a_{LA}^\dag)\ (a_L^\dag)^k\ \ket{00}\ket{00}_A,\\
    \ket{\psi}_2&=\frac1{\sqrt{2}}\text{sech}r\sum_{k=0}\frac{\lambda^k}{k!}\ (b_U^\dag)^k\ (b_{UA}^\dag+b_{LA}^\dag)\ (b_L^\dag)^k\ \ket{00}\ket{00}_A.\\
\end{aligned}
\end{equation}
Here, $b^\dag=\cos\theta\ a^\dag+\sin\theta\  a_A^\dag$ and $\ b_A^\dag=-\sin\theta\ a^\dag+\cos\theta\ a_A^\dag$. Note that $b^\dag$ denotes $b^\dag_U$ or $b^\dag_L$ while $b^\dag_A$ denotes $b^\dag_{UA}$ or $b^\dag_{LA}$.
\par
In principle, detecting $m'=0,n'=0$ leads to one photon addition to the TMSVS.  Substituting $b_L^\dag,\ b^\dag_U$ and $b^\dag_{UA},\ b^\dag_{LA}$ into the expression of $\ket{\psi}_2$, it follows that
\begin{equation} \label{eq:output_entangledsource}
\begin{aligned}
&\ket{\psi}_f=N\ _A\braket{00|\psi}_2 \\
&=N\frac{\text{sech}r}{\sqrt{2}}\sum_{k=0}\frac{\lambda^k}{k!}\ (\cos\theta a^\dag_U)^k\ (\cos\theta a^\dag_L)^k\ (-\sin\theta a^\dag_U-\sin\theta a^\dag_L) \ket{00}\\
&=N\frac{\text{sech}r}{\sqrt{2}}\sum_{k=0}(\lambda\cos^2\theta)^k\ (-\sin\theta a^\dag_U-\sin\theta a^\dag_L) \ket{kk}\\
&=-N\frac{\text{sech}r}{\sqrt{2}}\sum_{k=0}(\lambda\cos^2\theta)^k\sin\theta\sqrt{k+1}\ \bigg[\ket{k+1,k}+\ket{k,k+1}\bigg],
\end{aligned}
\end{equation}
where $N$ is the normalized factor.
\par
Up to an unimportant global phase, and for $\theta \neq 0$, the state at the output can be expressed as 
\begin{align}
\ket{\psi}_f=(1-\lambda^2\cos^4\theta)\sum_{k=0}\big[(\lambda\cos^2\theta)^k\sqrt{k+1}\big] \ket{\phi}_k,
\end{align}
where $\ket{\phi}_k=\frac1{\sqrt{2}}(\ket{k+1,k}+\ket{k,k+1})$. Alternatively, we can write the state as $\ket{\psi}_f=\sum_{k=0}\sqrt{p_k} \ket{\phi}_k$, in which $p_k$ is the probability of having the state $\ket{\phi}_k$ given by
$p_k=({1-\lambda^2\cos^4\theta})^2(\lambda\cos^2\theta)^{2k}({k+1})$. The state $\ket{\phi}_k$ is a maximally entangled state with any $k$. Therefore, the output state $\ket{\psi}_f$ reduces to superposition of different maximally entangled states with the probability determined by $p_k$. An interesting observation is to find which entangled state ($\ket{\phi}_k$) is most probable given the probability distribution $p_k$. To this end, one can easily find $p_{k+1}/p_k=(\lambda^2\cos^4\theta)(k+2)/(k+1)$. When the squeezing factor $r$ is small enough ($\lambda=\text{tanh}r$), we have $p_{k+1}/p_k<1$. Therefore, the most probable state can be attained for $k=0$, i.e., $\ket{\phi}_0=(\ket{1,0}+\ket{0,1})/\sqrt{2}$. However, for the case when $2\lambda^2\cos^4\theta>1$, one can determine the peak in the probability distribution by setting $p_{k+1}=p_k$. This gives $k=[1/(1-\lambda^2\cos^4\theta)]-2$.

\par
The output state for photon catalysis and photon subtraction can be obtained by applying $_A\bra{10}$ (or $_A\bra{01}$) and $_A\bra{11}$ to $\ket{\psi}_2$, respectively. 
\par
The success probability as well as the entanglement enhancement can be calculated from Eq. (\ref{eq:output_entangledsource}). It is worth to mention that, for $\ket{\psi}_2$ in the last step of Eq. (\ref{eq:output_entangledsource}), we apply Schmit decomposition to $\ket{\psi}_f$ to get the singular eigenvalue $C_k$, then use Eqs. (\ref{eq:logar2})-(\ref{eq:DeltaE}) to obtain the entanglement enhancement.
\begin{figure}[]
\begin{center}
\includegraphics[width=9cm]{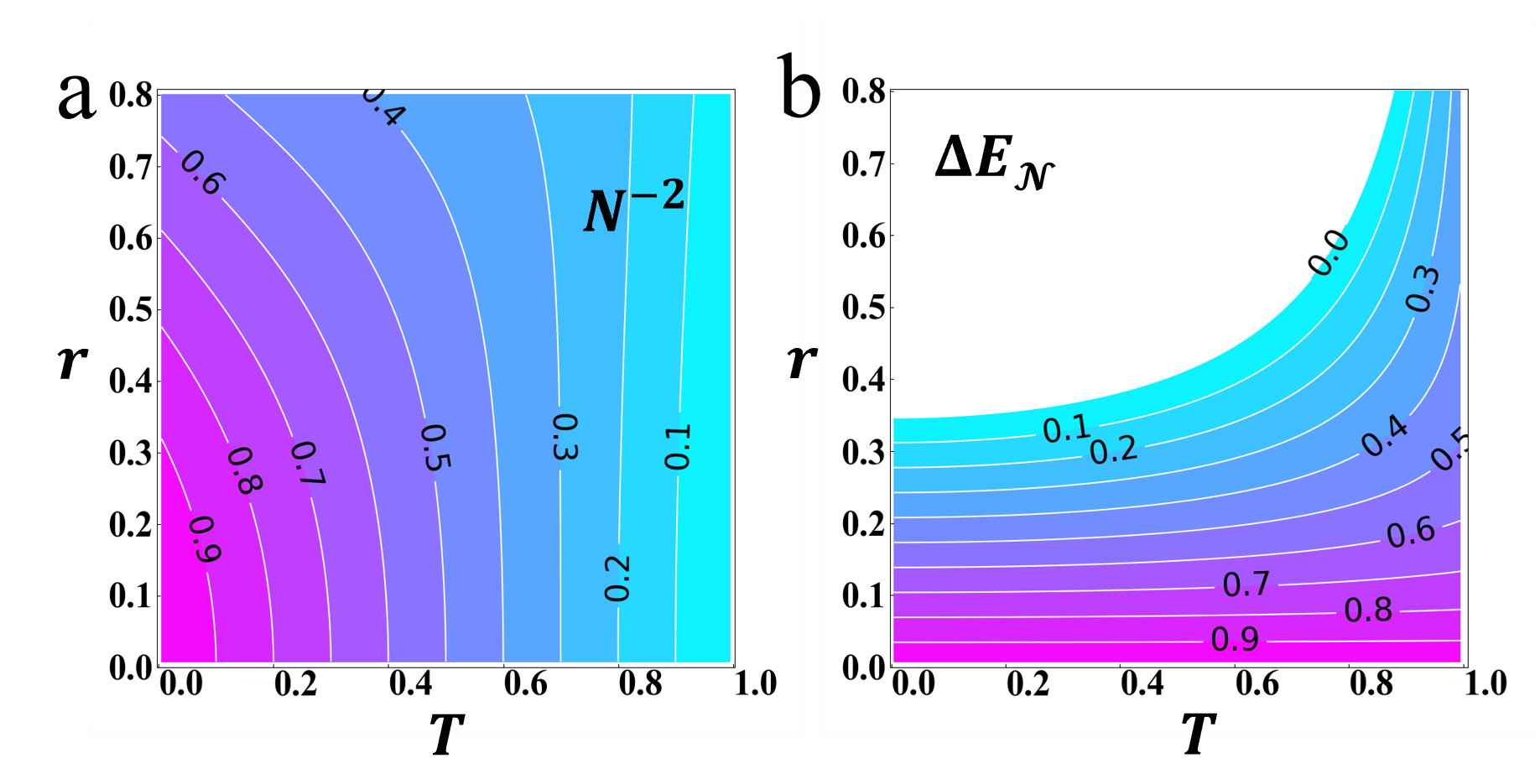}
\caption{Photon addition operation for detecting $\ket{00}_A$ in the protocol presented Fig. \ref{fig:threeBS}. In other words, $m=1,n=0$ and $m'=n'=0$. (a) is the success probability varying with squeezing factor $r$ and $T=\cos^2\theta$, while (b) is the enhancement entanglement.}
\label{fig:good}
\end{center}
\end{figure}
\par
\begin{figure}[]
\begin{center}
\includegraphics[width=9cm]{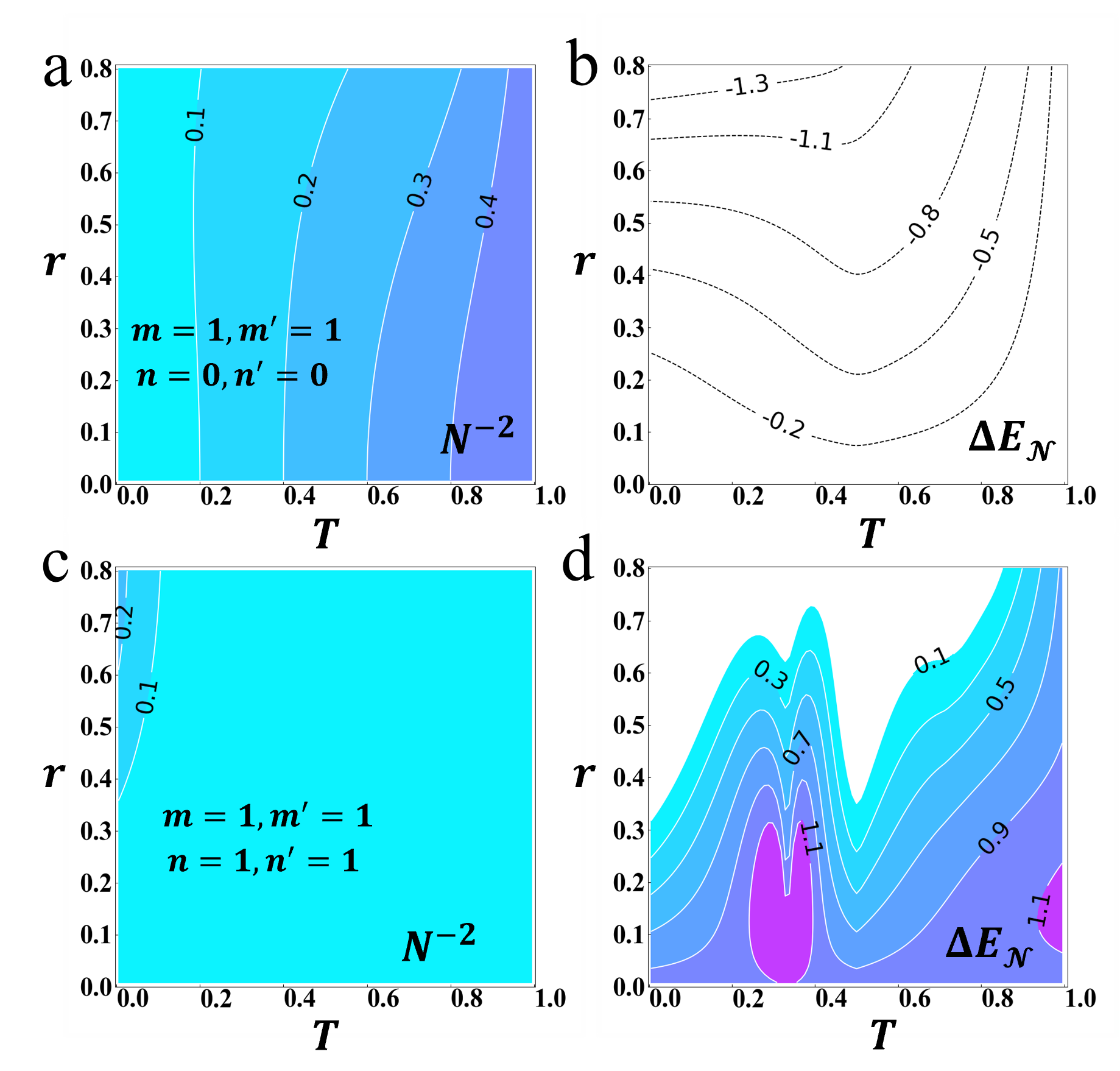}
\caption{ Photon catalysis for single photon auxiliary source ($m=1,n=0$) and conditionally detecting $m'=1,n'=0$, is  presented in (a) and (b). Photon catalysis for ancillary source $m=1,n=1$ and conditionally detecting $m'=1,n'=1$, is presented in (c) and (d). }
\label{fig:bad}
\end{center}
\end{figure}

We present the photon addition operation, for the protocol presented Fig. \ref{fig:threeBS}, in Fig. \ref{fig:good}. In this plot we have $m=1,n=0$ and $m'=n'=0$. Therefore, the detectors do not click in this case, i.e., the detected states is $\ket{00}_A$. Fig. \figref[(a)]{fig:good} presents the success probability varying with squeezing factor $r$ and $T=\cos^2\theta$; while Fig. \figref[(b)]{fig:good} shows the enhancement entanglement.

Compared to Fig. \ref{fig:three}, there is a considerable improvement in both success probability and entanglement with the same resources ($m=1,n=0$ and $m'=n'=0$). An interesting observation is that in the configuration of Fig. \ref{fig:threeBS}, photon catalysis does not provide a better result compared to photon addition. Instead, the single-photon addition with this process shows a much better outcome when compared to the settings from the previous setup. Therefore, the protocol in Fig. \ref{fig:threeBS} provides a better route for entanglement enhancing using beam-splitters and photon detection operations. This observation improves the previous efforts for enhancing the entanglement of TMSVS in Refs.\cite{hu2017continuous,sanaka2006filtering,resch2007entanglement}.

\par
Considering the setup in Fig. \ref{fig:two_BS}, the entanglement enhancement of the state by photon catalysis, obtained from the TMSVS, is shown to be higher than that of the photon-subtracted and photon-added state \cite{hu2017continuous}, which is in agreement with the observations of the current study. In multi-mode squeezing state, the superiority of photon subtraction to photon addition has been reported \cite{das2016superiority}. However, our investigations for TMSVS shows that photon addition performs better than both photon subtraction and photon catalysis.

\par
We also note that one might assume that the existence of the entanglement in the auxiliary modes, after $BS_A$, is the reason for better performance in this setting. However, this might not necessarily be the case.
To illustrate this, we consider the setting $m=1,n=1$, for which the entangled states after passing though $BS_A$ is  $\ket{\psi}_A=\frac{1}{\sqrt{2}}\ (-\ket{20}_A+\ket{02}_A)$, which is a maximally entangled state \cite{maleki2022distributed}. 
However, the result of such a process is not as good as what we can get in Fig. \ref{fig:threeBS}. Specific settings are presented in  Fig. \ref{fig:bad} for the illustration. 
As is shown in Figs. \figref[(a)]{fig:bad} and \figref[(c)]{fig:bad}, the success probabilities are much lower than the result in Fig. \figref[(a)]{fig:good}. Therefore, it is not proper to assume that the entanglement from the auxiliary modes is 'transmitted' to the TMSVS. As a result, single-photon addition through the setup is given in Fig. \ref{fig:threeBS} provides an optimal enhancement in the entanglement of a TMSVS. We, in fact, analysed several different setting beyond the scope of \ref{fig:bad}, however, non of the scenarios provide better result than what we presented in Fig. \ref{fig:good}. Therefore, the desirable method for enhancing entanglement seems to be the setting of Fig. \ref{fig:good}.

\par
An appealing feature of our optimal quantum entanglement in the second setup is that it does not require high-number Fock states. Basically, we only need single photon input state and dark detection in the detectors, as is analysed in Fig. \ref{fig:good}. Even, the scenario described in Fig. \ref{fig:bad} does not rely on high-number Fock states.
In Figs. \figref[(a)]{fig:bad} and \figref[(b)]{fig:bad}, for photon catalysis with $\ket{10}$ auxiliary state, the detected state is $\ket{10}$. Similarly, in Figs. \figref[(c)]{fig:bad} and \figref[(d)]{fig:bad}, for photon catalysis with $\ket{11}$ auxiliary state, the detecting state is $\ket{11}$. 
\par
The optimal situation, which requires single photon source as auxiliary input and detecting zero photons, provide another advantage from the photon detection point of view. Considering the fact that the dark counts of single photon avalanche detector (SPAD) can be as low as 50 per second while the weak photon signal reads as high as $10^4$ per second \cite{noh1991measurement,chen1999photon,zambra2005experimental}, ensuring that a no-photon event can  conveniently be distinguished from single-photon Fock states. Also, the detection of high-number Fock states are more challenging in general, while our proposed scenario does not rely on that. The conditional measurements can be done just with a singled photon detector, which simplifies the detection processing as well as improves the detection accuracy. 

Beside the photon detection error, the other imperfection may be accumulated from the BSs. Since the existing BSs are highly efficient in the realistic experiments, the error in BSs can be usually neglected. In fact, systems containing even a network of many BSs have intensively been considered in the literature. In this work, there is no need for high number of BSs, and thus, the imperfections cannot impair the performance of the protocols.


\section*{\centering\uppercase\expandafter{\romannumeral4}. Summary and conclusion}
In this work, we considered a protocol for entanglement enhancing in a TMSVS based on photon subtraction, photon catalysis, and photon addition. Central to such an operation is the task of mixing and detecting number states with each mode of TMSVS and non-deterministic detection. We analyzed various settings for the improvement of quantum entanglement and found an optimal setup for enhancing the entanglement of the state. 
To be more specific, we considered two different schemes for enhancing entanglement. In the first scheme each mode interacts with a Fock state in a beam-splitter. While, in the second scheme Fock states undergo a beam-splitter before interacting with the modes of TMSVS.

In the first scheme, photon catalysis outperforms both photon addition and subtraction operations. Nonetheless, the improvement and the success probability remain rather low even for photon catalysis. Of course, entanglement, in this case, can be improved more by using higher-number Fock states for resources. However, generating as well as detecting a high-number Fock state $\ket{m}$ is quite challenging. 

In contrary to the first scheme, it turns out that the optimal performance of the second setup can be achieved with simply using single photon sources. This, in fact, simplifies both photon generation and detection processes from the practical point of view. The optimal scenario presented in this work can drastically outperform all the settings of the first scheme, with more than three fold improvement of the entanglement.

Quantum entanglement improvement based on non-Gaussian operations are investigated from various perspectives, in recent years \cite{agarwal1991nonclassical,ourjoumtsev2009preparation,walschaers2017statistical,zavatta2004quantum,ourjoumtsev2007increasing,lvovsky2002quantum}. Such operations are shown to be useful for entanglement distillation \cite{lee2011enhancing,zhang2013continuous}, quantum commutation \cite{zavatta2009experimental,parigi2007probing} and quantum teleportation  \cite{opatrny2000improvement,walschaers2017statistical,hu2017continuous,ourjoumtsev2007increasing}. 
Therefore, the optimal entanglement enhancement proposed here can provide an important tool for such applications.
\par


\section*{Acknowledgments}
This research is supported by Project No. NPRP 13S-0205-200258 of the Qatar National Research Fund (QNRF). J.L is supported by the HEEP Graduate Fellowship.

\begin{widetext}

\section*{\centering Appendix: Derivation of $\hat{B}_{m,m'}$}

The BS operator $\hat{B}(\theta)$ performs a transformation to the input modes by
\begin{equation*} \label{A1}
\begin{aligned}
    {b_A}^\dag&=\hat{B}(\theta)\ \a^\dag\ \hat{B}(\theta)^\dag=\cos\theta\ \a^\dag+\sin\theta\  a^\dag, \\
    b^\dag&=\hat{B}(\theta)\ a^\dag\ \hat{B}(\theta)^\dag=-\sin\theta\ \a^\dag+\cos\theta\ a^\dag.
\end{aligned} \tag{A1}   
\end{equation*}

From Eq. (\ref{eq:Bmm'}), in order to determine the output state we need to compute $\hat{B}_{m,m'}$. 
\begin{equation*} \label{A2}
\begin{aligned}
\hat{B}_{m,m'} &= _A\bra{m'}\hat{B}(\theta)\ket{m}_A = _A\bra{m'}\hat{B}(\theta)\sum_{k=0}\ket{m}_A\ket{k}\bra{k} \\
&=_A\bra{m'}\sum_{k=0}\hat{B}(\theta)(\a^\dag)^m(a^\dag)^k\frac1{\sqrt{m!}}\frac1{\sqrt{k!}}\ket{0}_A\ket{0}\bra{k} \\
&=_A\bra{m'}\sum_{k=0}(\b^\dag)^m(b^\dag)^k\frac1{\sqrt{m!}}\frac1{\sqrt{k!}}\ket{0}_A\ket{0}\bra{k} \\
&=\sum_{k=0}B_{m,m',k}\ket{k+m-m'}\bra{k},
\end{aligned} \tag{A2}
\end{equation*}

Considering $_A\bra{m'}$, that is inserted from the measurement, the non-zero term must contain $(\a^\dag)^{m'}\ket{0}$. Substituting the expressions for $\b^\dag$ and $b^\dag$ from Eq. (\ref{A1}) in Eq. (\ref{A2}) we have
\begin{equation*} \label{A3}
\begin{aligned}
\hat{B}_{m,m'} &=_A\bra{m'}\sum_{k=0}(\cos\theta\ \a^\dag+\sin\theta\  a^\dag)^m(-\sin\theta\ \a^\dag+\cos\theta\ a^\dag)^k\frac1{\sqrt{m!}}\frac1{\sqrt{k!}}\ket{0}_A\ket{0}\bra{k} \\
&=_A\bra{m'}\sum_{k=0}\sum_{i=0}^{m'}\binom{m}{i}(\cos\theta)^i(\sin\theta)^{m-i}\binom{k}{m'-i}(-\sin\theta)^{m'-i}(\cos\theta)^{k+i-m'}(\a^\dag)^{m'}(a^\dag)^{m+k-m'}\frac1{\sqrt{m!}}\frac1{\sqrt{k!}}\ket{0}_A\ket{0}\bra{k}\\
&=\sum_{k=0}\sum_{i=0}^{m'}(-1)^{m'-i}\binom{m}{i}\binom{k}{m'-i}(\cos\theta)^{k+2i-m'}(\sin\theta)^{m+m'-2i}\frac{\sqrt{m'!}}{\sqrt{m!}}\frac{\sqrt{(k+m-m')!}}{\sqrt{k!}}\ket{k+m-m'}\bra{k}\\
&=\sum_{k=0}B_{m,m',k}\ket{k+m-m'}\bra{k}, \\
\text{with}\quad B_{m,m',k}&=\sum_{i=0}^{m'}(-1)^{m'-i}\binom{m}{i}\binom{k}{m'-i}(\cos\theta)^{k+2i-m'}(\sin\theta)^{m+m'-2i}\frac{\sqrt{m'!}}{\sqrt{m!}}\frac{\sqrt{(k+m-m')!}}{\sqrt{k!}}.
\end{aligned} \tag{A3}
\end{equation*}
Note that $\binom{a}{b}=0$ if $a<b$, thus, the situation when $m'>m$ or $m'<m, m'>k$ is included in Eq. (\ref{A3}).

\end{widetext}

\bibliography{ref}
\end{document}